\documentclass[final,12pt]{elsarticle}

\usepackage{color}
\usepackage{amssymb}
\usepackage{epsfig}
\usepackage{epstopdf}
\usepackage{graphicx,color}
\usepackage[usenames,dvipsnames]{xcolor}
\usepackage[section]{placeins}
\usepackage{amsmath}
\usepackage[normalem]{ulem}
\usepackage{booktabs}

\usepackage{xcolor}
\colorlet{darkgreen}{green!50!black}
\colorlet{orange}{red!50!yellow}
\colorlet{darkblue}{blue!60!black}
\colorlet{darkred}{red!80!black}

\usepackage{mathtools}
\usepackage{mathrsfs}
\usepackage{bm}
\usepackage{indentfirst}

\usepackage{xcolor}

\journal{Physics Letters B}
\begin{document}
\begin{frontmatter}
\title{ 
Color-suppression of non-planar diagrams in  bosonic bound states}
\author{J.~H.~Alvarenga Nogueira$^{a,b}$}
\author{Chueng-Ryong Ji$^c$}
\author{E.~Ydrefors$^a$} 
\author{T.~Frederico$^a$}
\address{$^a$Instituto Tecnol\'ogico de Aeron\'autica, DCTA, 
12228-900 S\~ao Jos\'e dos Campos,~Brazil}
\address{$^b$Dipartimento di Fisica, Universit\`a di Roma ``La Sapienza" \\
INFN, Sezione di Roma, 
Piazzale A. Moro 5 - 00187 Roma, Italy}
\address{$^c$Department of Physics, North Carolina State University, Raleigh, North Carolina, 27695-8202, USA}
\date{\today}

\begin{abstract}
We study the suppression of non-planar diagrams in a scalar QCD model of a meson system in $3+1$  space-time 
dimensions due to the inclusion of the color degrees of freedom. As a prototype of the color-singlet meson, we consider a 
flavor-nonsinglet system consisting of a scalar-quark and a scalar-antiquark with equal masses exchanging a scalar-gluon 
of  a different mass, which is investigated within the framework of the homogeneous Bethe-Salpeter equation. 
The equation is solved by using the Nakanishi representation for the manifestly covariant bound-state amplitude 
and its light-front projection. The resulting non-singular integral equation is solved numerically. The damping of
the impact of the cross-ladder kernel on the binding energies are studied in detail. 
The color-suppression of the cross-ladder effects on the light-front wave function and the elastic electromagnetic 
form factor are also discussed.
As our results show, the suppression appears significantly large for $N_c=3$, which supports the use of rainbow-ladder truncations 
in  practical nonperturbative calculations within QCD.
\end{abstract}
\end{frontmatter}

\section{Introduction}\label{intr}

Relativistic studies of non-perburbative systems are essential in order to understand the properties of composite systems, 
which are characterized by strong coupling constant and/or small constituent masses~\cite{Ji94}. One relevant example is the calculation 
of masses and form factors of light mesons, consisting of quarks with small masses and gluons.  
The light-front two-body bound-state equation\cite{Ji94} was extended to the full light-front dynamic kernel including the ladder, cross-ladder, 
stretched-box, and particle-antiparticle creation/annihilation effects to study the contributions of higher Fock states~\cite{Ji-Tokunaga12}
and its link to the powerful numerical approach known as the Feynman-Schwinger representation (FSR) approach~\cite{Savkli, Tjon} 
as well as the manifestly covariant Bethe-Salpeter formulation~\cite{BSE} has been discussed extensively.

An analytically tractable tool for the bound-state problem, what is also known as the most orthodox tool for dealing with 
the relativistic two-body problem in quantum field theory, is the Bethe-Salpeter equation (BSE), proposed in the 1950s \cite{BSE},
utilizing the Green's functions of covariant perturbation theory (see also \cite{Sales:1999ec}).
This equation has for a long time been solved by a transformation 
to Euclidean space, by using an analytical continuation to the complex-energy plane \cite{Wick54}. Solutions in Minkowski 
space are however essential to perform calculations of structure-dependent observables such as form factors both in spacelike and 
timelike regions. One of the first solutions of the BSE was obtained by Kusaka and Williams in Ref.~\cite{kusaka}, who studied 
the bound-state system of two scalar bosons interacting through the exchange of another boson, by  using the Nakanishi integral 
representation (NIR, \cite{nakanishi}) and the ladder kernel in the lowest approximation.
Another major improvement regarding the treatment of the Bethe-Salpeter equation in Minkowski space was 
performed by Karmanov and Carbonell in Ref.~\cite{karmanov}. They studied the two-boson bound-state system in 
the ladder approximation, by using the NIR and the explicitly covariant Light-Front formalism \cite{Carbonell98}. 
They reformulated the BSE in a more convenient form and the ladder plus cross-ladder kernel was subsequently considered in Ref.~\cite{cross-box}. 

In \cite{Gigante17} Gigante et al studied the impact of the cross-ladder contribution on the bound-state structure of the two-boson 
system in great detail. The elastic electromagnetic form factor was also computed by considering the two-current contribution 
in addition to the one given by the impulse approximation. It was found that the cross-ladder contribution to the kernel 
has a rather large impact on the coupling constants as well as the asymptotic behavior of the light-front wave functions, 
i.e.~the coupling constant changed by about 30-50\% depending on the considered binding energy. 

Differently from this theory, 
in  quantum chromodynamics (QCD) one has also to consider the color degree of freedom. It was shown  in \cite{thooft-planar} 
that the non-planar diagrams, like e.g. the cross-ladder, have a vanishing contribution in the limit $N_c\rightarrow \infty$, 
where $N_c$ is the number of colors. 
$SU(N)_c$ theories of QCD in 1+1 dimensions have been extensively studied by 't Hooft \cite{thooft} and also 
by Hornbostel et al \cite{Hornbostel90}. 

In this work, we consider a scalar QCD model to study a prototype of a flavor nonsinglet meson system of a scalar-quark and a scalar-antiquark 
with equal masses exchanging a scalar-gluon of different mass in  3+1 dimensions. 
This is obtained by adding the appropriate 
color factors in the manifestly covariant BSE and its light-front projection. We study quantitatively the suppression of the 
contribution coming from the cross-ladder interaction kernel in the nonperturbative problem of the color singlet two-boson bound state
mass and structure, due to the color factors in the kernel
for different number of colors $N_c=N$ with $N = 2, 3, 4$. The effects of the non-planar cross-ladder interaction kernel on the coupling constant for a 
given bound state mass as well  as in light-front wave function, and the elastic electromagnetic form factor are studied in detail.

This work is organized as follows. In Sec.~\ref{Sec:Theory} we briefly introduce the Bethe-Salpeter equation and the Nakanishi integral representation. 
The formalism for introducing the color factors is also presented. Then, in Sec.~\ref{Sec:Results} we discuss our numerical results for the coupling 
constants for a given bound state mass, the light-front wave function and the elastic electromagnetic form factor. 
Finally, in Sec.~\ref{Sec:Summary and Outlook} we summarize our work and give an outlook.

\section{Theoretical framework \label{Sec:Theory}} 

\subsection{Bethe-Salpeter equation}
The Bethe-Salpeter equation (BSE) in Minkowski space, for two spinless particles,
reads:
\begin{eqnarray}\label{bs0}
\Phi(k,p)=S(\eta_1 p + k) S(\eta_2 p - k)
\int \frac{d^4k'}{(2\pi)^4}iK(k,k',p)\Phi(k',p),
\end{eqnarray}
where the propagators $S(p')$ are in general dressed and can be represented by the K\"allen-Lehmann spectral representation as 
\begin{equation}
S(p') = \int_0^\infty d \gamma \frac{\rho(\gamma)}{p'^2 - \gamma + i \epsilon}\, ,
\end{equation}
which raise up the following
\begin{eqnarray}\label{bs1}
\Phi(k,p)= \int_0^\infty d \gamma \frac{\rho(\gamma)}{(\eta_1 p+k)^2 - \gamma + i \epsilon} 
\int_0^\infty d \gamma' \frac{\rho(\gamma')}{(\eta_2 p-k)^2 - \gamma' + i \epsilon}
\int \frac{d^4k'}{(2\pi)^4}iK(k,k',p)\Phi(k',p),
\end{eqnarray}
where in the interaction kernel, $K(k,k',p)$, the exchanged boson is in general also dressed. For a prototype of 
flavor nonsinglet meson system, $K(k,k',p)$ does not get complicated by the annihilation 
 process, but it keeps the usual ladder and cross-ladder irreducible kernels, 
which have been extensively discussed in Refs. \cite{Ji-Tokunaga12} and \cite{Gigante17}. 
The idea of our paper is to start exploring the effects of color degrees of freedom on the non-planar diagrams in the
simplest possible way, without considering dressed propagators. For that purpose we put $\rho(\gamma) = \delta (\gamma - m^2)$, 
which gives
\begin{eqnarray}\label{bs}
\Phi(k,p)=\frac{i^2}{\left[(\frac{p}{2}+k)^2-m^2+i\epsilon\right]
\left[(\frac{p}{2}-k)^2-m^2+i\epsilon\right]}
\int \frac{d^4k'}{(2\pi)^4}iK(k,k',p)\Phi(k',p),
\end{eqnarray}
where the interaction kernel $K$ is given by a sum of irreducible Feynman
diagrams and considering the equal partition for the momentum fraction $\eta_1=\eta_2=1/2$. The ladder kernel  is considered in most of works,
but here we also incorporate the cross-ladder contribution. 

The Bethe-Salpeter (BS) amplitude can be found in the form of the Nakanishi integral representation~\cite{kusaka,nakanishi}:
\begin{eqnarray}\label{bsint}
\Phi(k,p)=-{i}\int_{-1}^1dz\int_0^{\infty}d\gamma 
\frac{g(\gamma,z)}{\left[\gamma+m^2
-\frac{1}{4}M^2-k^2-p\cdot k\; z-i\epsilon\right]^3}.
\end{eqnarray}
The weight function $g(\gamma,z)$ is non-singular, whereas
the singularities of the BS amplitude are fully reproduced by this
integral. The BS amplitude in the form (\ref{bsint}) is then substituted into the BS
equation (\ref{bs}) and after some mathematical transformations, namely upon integration in $k^-=k^0-k^3$ on both sides of the BS equation, 
one obtains the following non-singular integral equation for
$g(\gamma,z)$ \cite{Gigante17}:
\begin{eqnarray} \label{bsnew}
\int_0^{\infty}\frac{g(\gamma',z)d\gamma'}{\Bigl[\gamma'+\gamma
+z^2 m^2+(1-z^2)\kappa^2\Bigr]^2} =
\int_0^{\infty}d\gamma'\int_{-1}^{1}dz'\;V(\gamma,z,\gamma',z')
g(\gamma',z'),
\end{eqnarray}
where for bound states $\kappa^2 = m^2- \frac{1}{4}M^2 > 0$ and $V$ is expressed in terms of the kernel $K$~\cite{cross-box}.

Furthermore, the s-wave valence light-front wave function can be computed from
\begin{equation}
\Psi_{\text{LF}}(\gamma,z)=\frac{1-z^2}{4}\int_0^{\infty}\frac{g(\gamma',z)d\gamma'}{[\gamma'+\gamma+z^2m^2+(1-z^2)\kappa^2]^2},
\label{Eq:lfwfn}
\end{equation}
derived in Ref. \cite{karmanov}.

\subsection{Scalar QCD}

\begin{figure}
\begin{center}
\includegraphics[scale=.6]{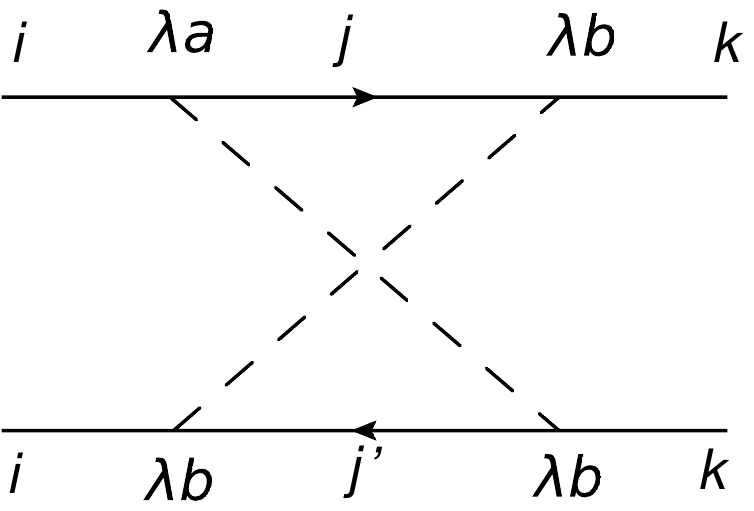}
\caption{Cross-ladder graph with color matrices.}\label{crossed}
\end{center}
\end{figure}

The extension of the BSE to a scalar QCD model is obtained simply by introducing  the color matrices in the interaction vertices appearing in the 
kernel diagrams. The Feynman rules are first used and then the kernel dependence on $N$ is derived by performing the trace of the product of 
Gell-Mann matrices, corresponding to a colorless composite two-boson system. For the ladder kernel this computation is straightforward:
\begin{eqnarray}
tr[(\lambda^a)_{ji} (\lambda^a)_{ij}] &=& \sum_a (\lambda^a)_{ji} (\lambda^a)_{ij} = \frac{1}{2} \sum_{i,j=1}^3 \left( \delta_{jj} \delta_{ii} - \frac{1}{N} \delta_{ji} \delta_{ij} \right) \nonumber \\
&=& \frac{1}{2} \left(N^2 - \frac{1}{N} \sum_{i=1}^3 \delta_{ii} \right) = \frac{N^2-1}{2},
\end{eqnarray}
where the internal boson line factors have been replaced by the corresponding $SU(N)$ projection operators (see Ref.~\cite{cvitanovic}).
As can be seen in Fig.~\ref{crossed}, the color factors are given by $(\lambda_a)_{ij}(\lambda_a)_{k j'}$ and $(\lambda_b)_{jk}(\lambda_b)_{j' i}$, and they read 
\begin{equation}
\sum_a (\lambda_a)_{ji}(\lambda_a)_{j' k} = \frac{1}{2} \left( \delta_{j k} \delta_{i j'} - \frac{1}{N} \delta_{i j} \delta_{j' k} \right).
\label{Eq:color_factors}
\end{equation}
By using Eq.~\eqref{Eq:color_factors} and subsequently performing the relevant traces one obtains for the cross-ladder kernel that
\begin{equation}
tr[\lambda^a \lambda^b \lambda^a \lambda^b]=- \frac{(N^2-1)}{4N}.
\end{equation}

\begin{figure}
\begin{center}
\includegraphics[scale=.75]{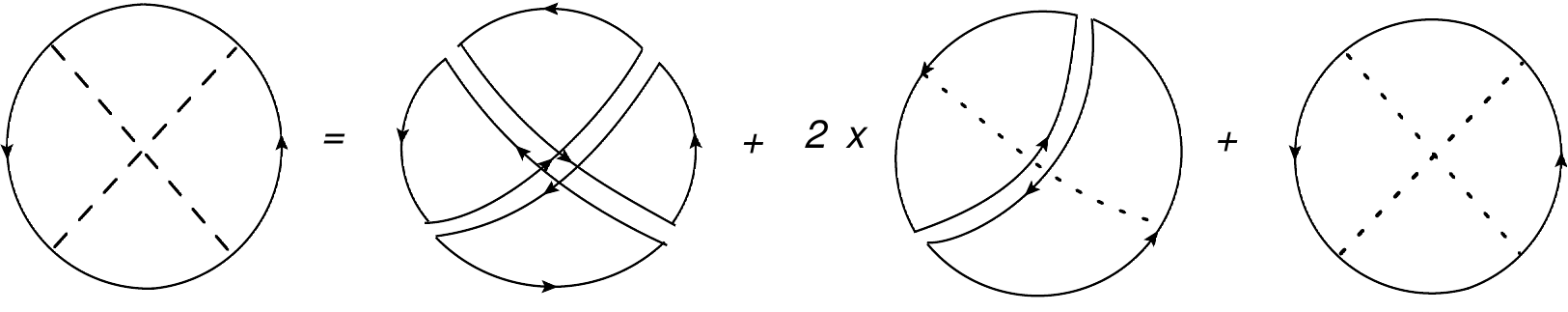}
\caption{Diagrams illustrating the color flow in the cross-ladder graph.}\label{crossed-color-flow}
\end{center}
\end{figure}

Consequently, the cross-ladder kernel is suppressed when compared to the $N^2-1$ factor obtained for the ladder one. In addition, we observe the
change in sign of the cross-ladder kernel, which turns out to be repulsive in the present model.
The relative damping coming from the color factor is also confirmed by Fig.~\ref{crossed-color-flow} where the color flow diagrams are shown. 
In the figure each loop is associated with a factor of $N$ and each dotted line, which corresponds to a phantom propagator, gives a factor $-1/N$. 
A similar analysis for the planar two-boson exchange kernel leads to a factor
 $\frac{N}{4}\left(N-\frac{1}{N}\right)^2=\frac{1}{4} \left( N^3-2N + \frac{1}{N} \right)$, which shows that the non-planar graphs are
  much more suppressed than the planar ones. This was also showed by 't Hooft for a $1+1$ model of mesons in the light-front form~\cite{thooft}.

\section{Results \label{Sec:Results}}

In the present work we solved the generalized integral equation \eqref{bsnew} for a fixed binding energy, $B$, and  
exchanged boson mass, $\mu$, by using the same method as in \cite{Gigante17}. That is, a basis expansion in terms of Laguerre polynomials 
was used for the noncompact variable and Gegenbauer polynomials for the compact one. For more details, see also Ref.~\cite{Frederico14}. 
The Eq.~\eqref{bsnew} is then transformed into a generalized eigenvalue problem, which is solved for the coupling constant $g^2$ 
and the coefficients of the basis functions that determine the Nakanishi weight $g(\gamma,z)$. 

\begin{figure}[!hbt]
\begin{center}
\includegraphics[scale=0.4]{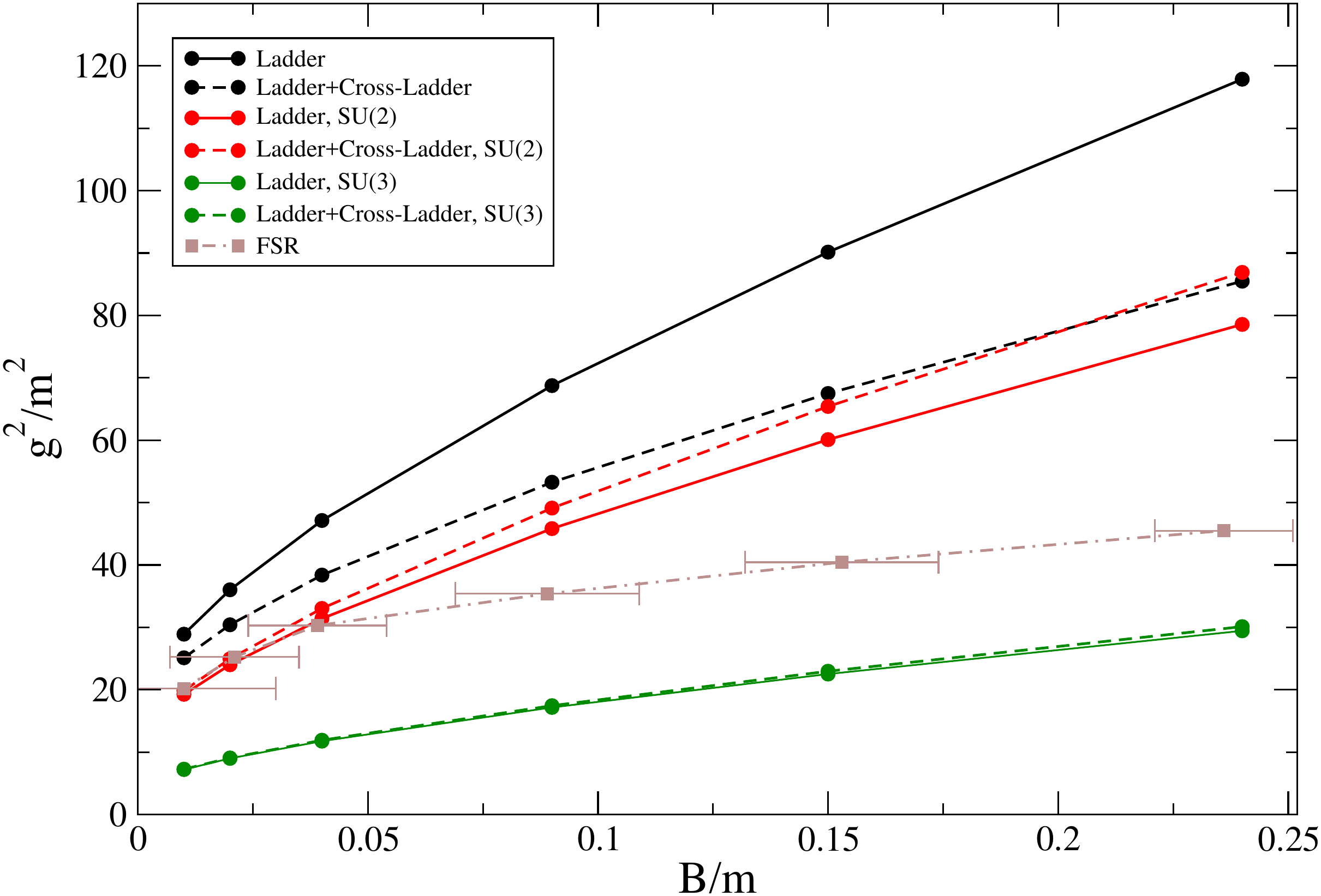}
\caption{Coupling constant for different values of the binding energy $B$ obtained by the Bethe-Salpeter ladder (L) and ladder plus cross-ladder kernel (L + CL), for an exchanged mass of $\mu=0.15\,m$. The figure shows the results computed without color factors and the corresponding 
ones for $N=2$ and 3. Our results  are also compared with the Feynman-Schwinger representation calculations from Ref.~\cite{Tjon}.}\label{Bvsg2-BSE-FSR}
\end{center}
\end{figure}

In Fig.~\ref{Bvsg2-BSE-FSR}, we show the coupling constant for an exchanged mass of $\mu=0.15\,m$ and different values of the binding energy, $B$, 
computed by solving the BSE for the ladder (L) and ladder plus cross-ladder (CL) kernels. In the figure these results are compared with the 
ones obtained by adding the color factors for $N=2$ and 3, and also with the Feynman-Schwinger representation (FSR) calculations of \cite{Tjon}. In the FSR calculation all the crossed combinations are considered without color degree of freedom. 
As discussed before, the contributions from the non-planar graphs should be suppressed when color is included, which could change completely the results 
of the FSR calculations approaching them to the ladder case. In the present case, the discrepancy between our results (including color) and the FSR ones is presumably  due to the absence of color factors in the latter case.

From the results we may conclude that without the color factors, the cross-ladder contribution is attractive, and quite important. On the contrary, 
with color factors it is repulsive and suppressed. For example, for $B=1.5\,m$ the cross-ladder contribution is about 52\% (no color factor). 
Then, for $N=2$ and $N=3$ 
it is respectively, $15\%$ and $3\%$, and with $N=4$  the effect is only $1\%$ (see Table \ref{Tab:g2}).  Additionally, it is seen that  $g^2$, for a given $B$, decreases when the number of colors  increases. This is in agreement with the 't Hooft's limit which states that $g^2 N$ should become a constant as 
  $N\rightarrow\infty$.
%
We note that we obtained the results also in 2+1 dimensions but the essential features of results in 3+1 dimensions are unaltered in 2+1 dimensions. Thus, we do not present the results in 2+1 dimensions in this work although the study of the kernel truncation on 2+1 systems turns out to be interesting since the BSE is widely applied to study 2D materials within condensed matter physics.
\vspace{1cm}

\begin{table}[!htb]
\begin{center}
\begin{tabular}{c c c c c c}
\toprule
$B/m$ & $\mu/m$ & $g^2_\text{L}/g^2_\text{L+CL}$ & $g^2_\text{L}/g^2_\text{L+CL}$ & $g^2_\text{L}/g^2_\text{L+CL}$ & $g^2_\text{L}/g^2_\text{L+CL}$   \\
& & {\footnotesize(BSE)} & {\footnotesize({\textit N}=2)} & {\footnotesize({\textit N}=3)} & {\footnotesize({\textit N}=4)} \\
\midrule
0.1 & 0.001 & 1.3181 & 0.9246 & 0.9823 &  0.9930\\
0.1 & 0.150 & 1.2998 & 0.9303 & 0.9835 &  0.9935	\\
1.5 & 0.001 & 1.5199 & 0.8456 & 0.9661 &  0.9867\\
1.5 & 0.150 & 1.5174 & 0.8467 & 0.9663 &  0.9868\\
\bottomrule
\end{tabular}
\end{center}
\caption{Ratios of the coupling constants, $g^2$, calculated with the ladder (L) and ladder plus cross-ladder (L+CL) kernels, corresponding to the binding energies $B=0.1\,m$ and $B=1.5\,m$, and for and exchanged mass of $\mu=0.001\,m$ and $\mu=0.15\,m$ respectively. In the table the results for the bare BSE is compared with the ones computed with $N=2, 3, 4$ colors. \label{Tab:g2}}
\end{table}

It is worthwhile to study how the suppression changes in the limit where the scalar QCD gluon mass goes to zero 
in order to verify the robustness of our calculations when the scalar exchanged mass is changed between 
the perturbative value to a finite one, reminiscent of the running gluon mass,
as the model approaches to the more realistic QCD scenario.
In Table \ref{Tab:g2} the ratios of the coupling constants, $g^2$, corresponding to the ladder and ladder plus cross-ladder kernels, are displayed for two different values of the exchanged mass, namely $\mu=0.001\,m$ and $\mu=0.15\,m$,  both for a weakly bound system ($B=0.1\,m$) and a strongly bound one ($B=1.5\,m$). 
In the table the results for the bare BSE are compared with the ones corresponding to $N=2, 3$ and 4 colors.

\begin{figure}[!hbt]
\begin{center}
\includegraphics[scale=.4]{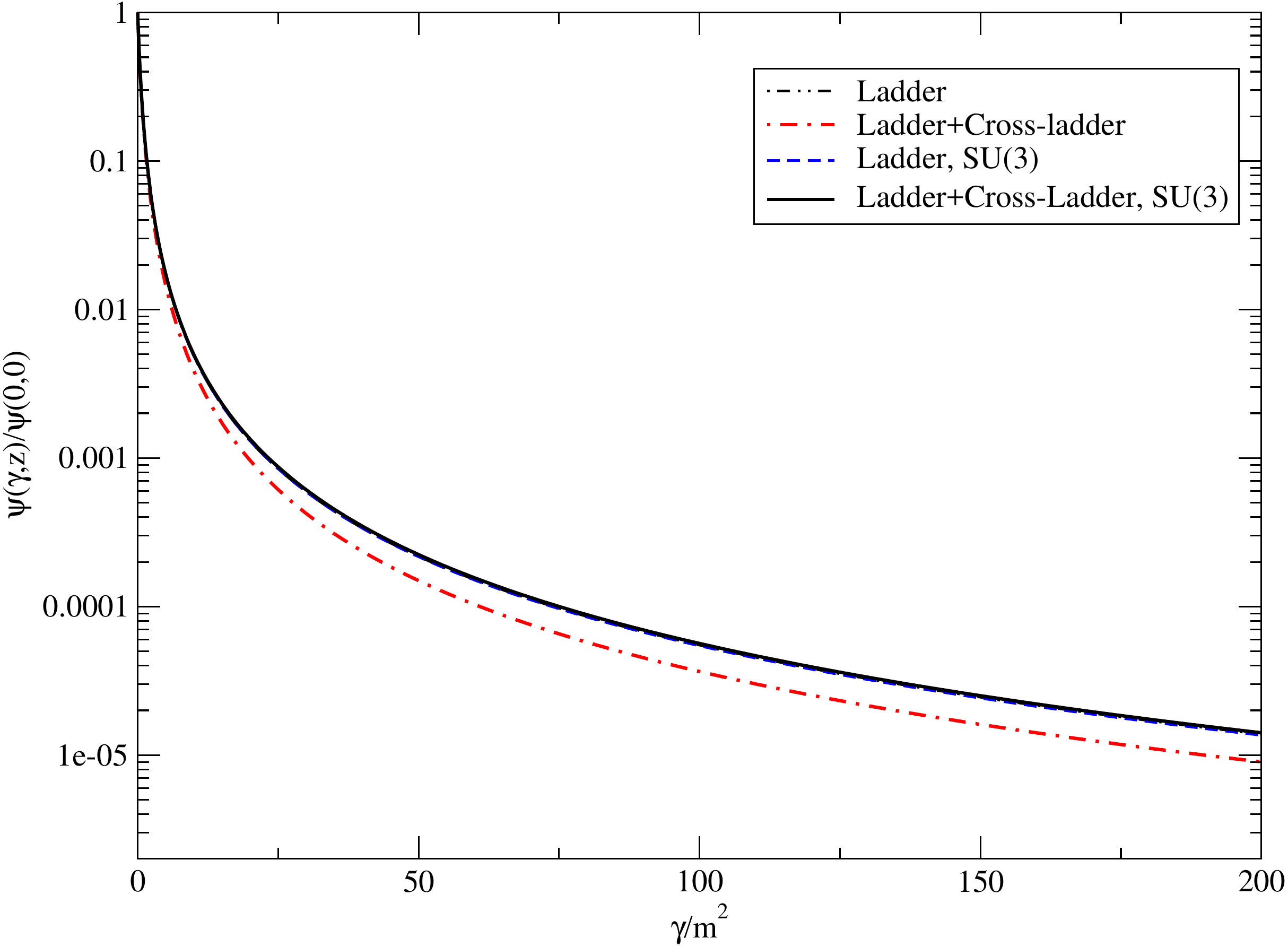} 
\caption{Valence wave function as a function of $\gamma=k_\perp^2$ computed with the Bethe-Salpeter amplitude for ladder and ladder plus cross-ladder kernels. The wave function obtained with $N=3$  is compared with the one without color factors, for a small exchanged mass of $\mu=0.001\,m$. Used values of the other input parameters: $B=1.0\,m$ and $z=0.0$.}\label{wavefunction}
\end{center}
\end{figure}

\begin{figure}[!hbt]
\begin{center}
\includegraphics[scale=.4]{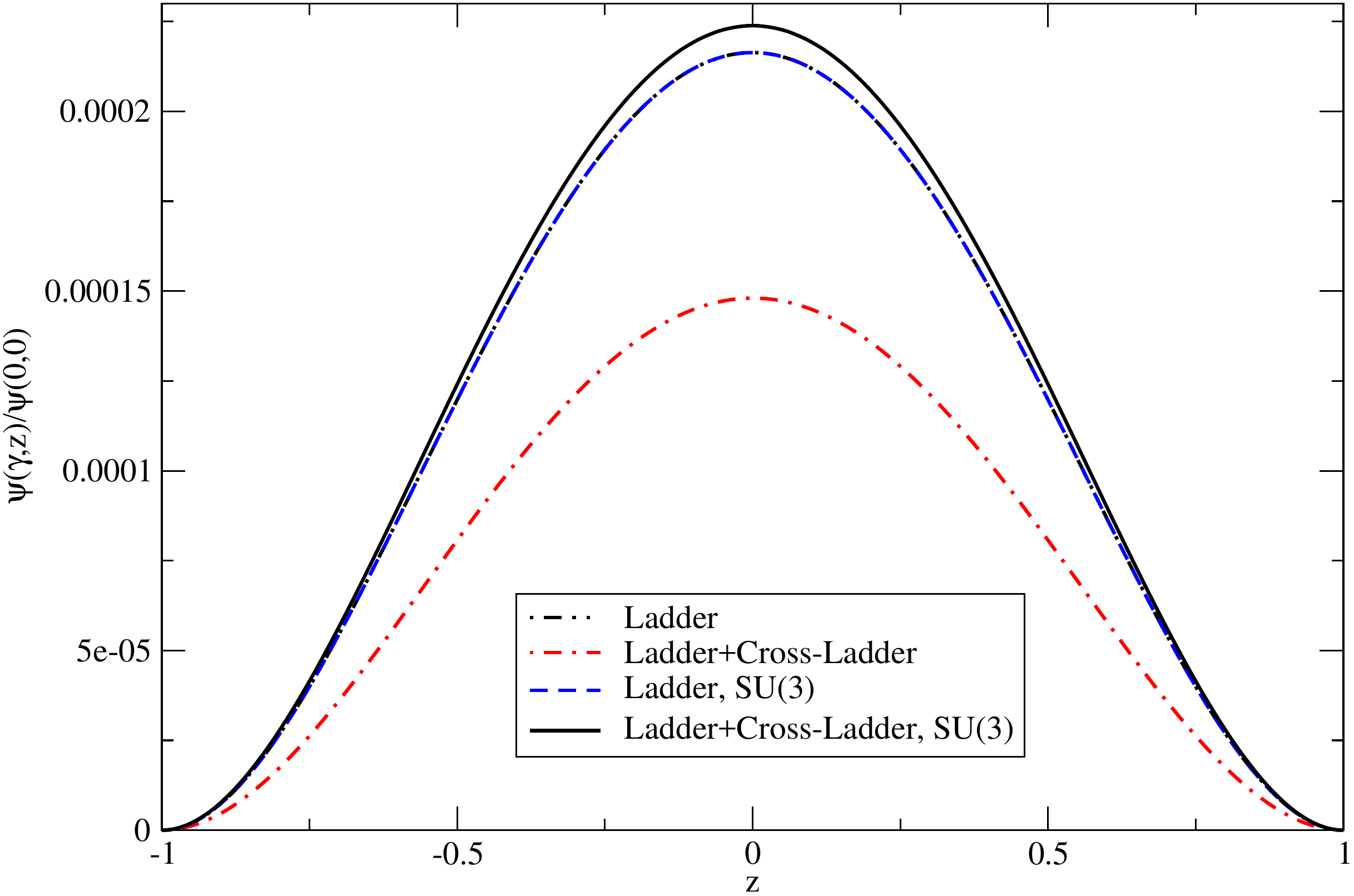} 
\caption{Valence wave function as a function of $z$ computed with the Bethe-Salpeter amplitude using the ladder and ladder plus cross-ladder kernels, for $\mu=0.001\,m$. In the figure the results for $N=3$ are compared with the ones without color factors. Adopted values of the other input parameters: $B=1.0\,m$, $\gamma=50\,m^2$.}\label{wavefunction-z}
\end{center}
\end{figure}

\begin{figure}[!hbt]
\begin{center}
\includegraphics[scale=.32]{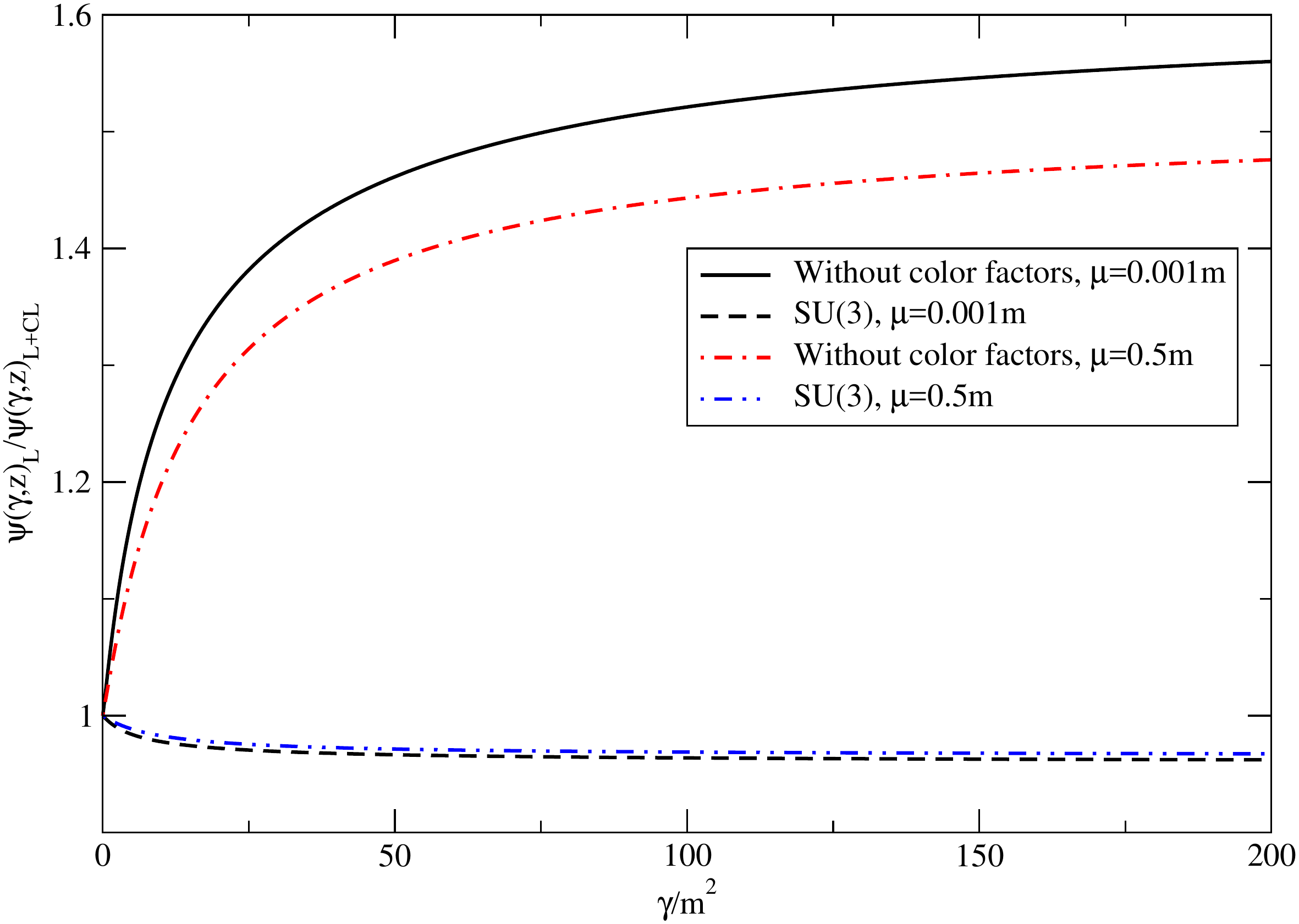} \quad \includegraphics[scale=.32]{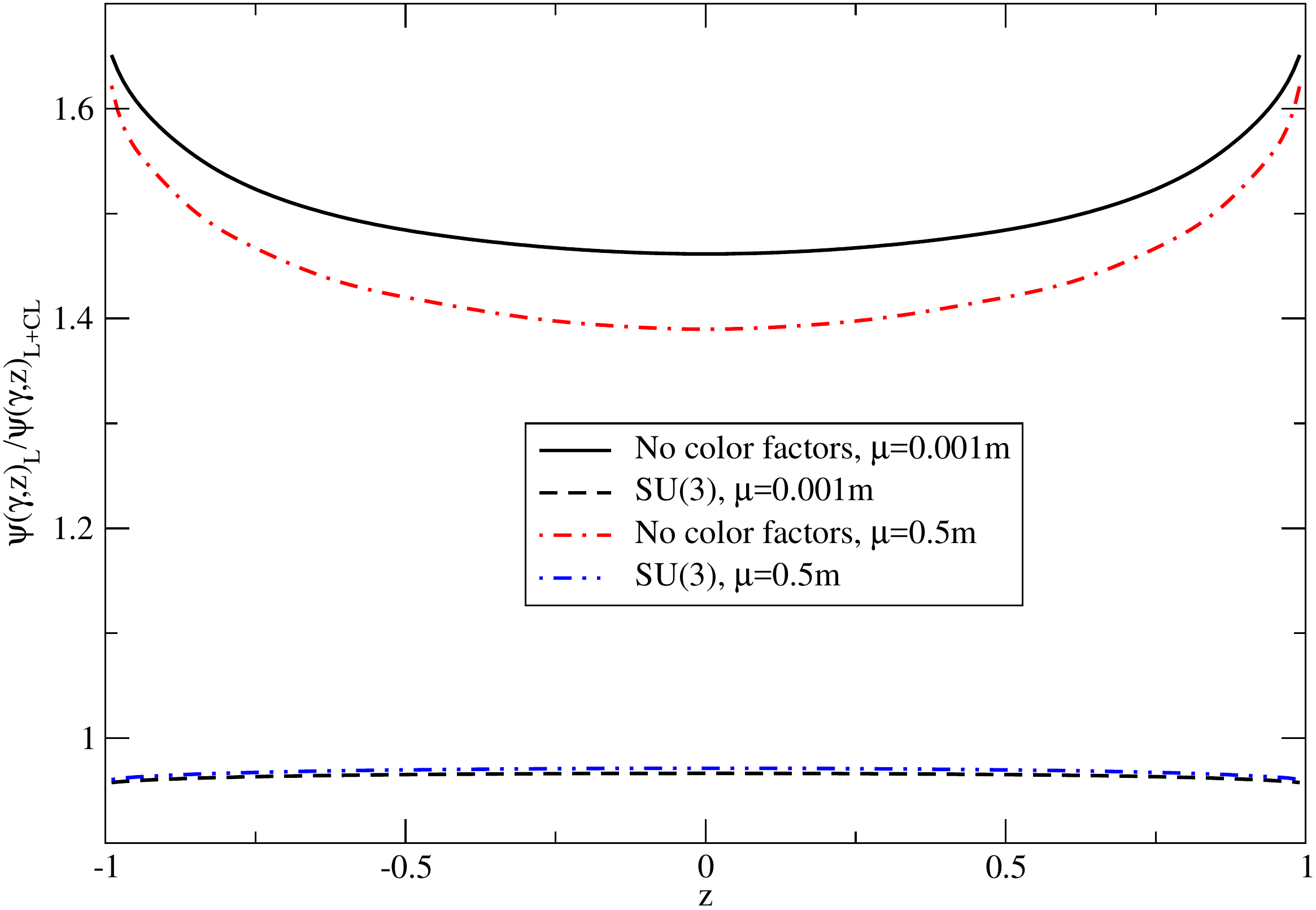} 
\caption{Ratio between the valence wave functions calculated with the ladder and ladder plus cross ladder kernels, for an exchanged mass of $\mu=0.001\,m$ and $\mu=0.5\,m$ respectively. In the figures the results for $N=3$  are compared with the ones obtained without color factors. The left panel shows the behavior with respect to $\gamma$ for $z=0$. Similarly, in the right panel the wave function is shown as a function of $z$ for $\gamma=50\,m^2$. The results correspond to a binding energy of $B=1.0\,m$.}\label{wavefunction-ratios}
\end{center}
\end{figure}

Additionally, the valence light-front wave function for fixed $B$ and $\mu$ was computed in the present study by using Eq.~\eqref{Eq:lfwfn}. In 
Fig.~\ref{wavefunction} we show the dependence of the wave function with respect to the variable $\gamma=k_\perp^2$ for $B=1.0\,m$, and the exchanged mass $\mu=0.001\,m$  and with fixed $z=0.0$. 
In the calculations, we normalize the LF wave function to be 1 at the point $(z,\gamma)=(0,0)$. Although the ladder result for SU(3) and the corresponding ladder result without the color degrees of freedom differ by the constant given by the color factor, this constant difference is lost when we normalize the LF wave function, and thus, the two ladder results are on top of each other in Fig.~\ref{wavefunction}.
Similarly, in Fig.~\ref{wavefunction-z} where the dependence of the wave function on $z$ for $\gamma=50.0\,m^2$ is displayed, 
the two ladder results with and without the SU(3) color factor are on top of each other. In Fig.~\ref{wavefunction-z},  we compare the BSE results (with no color factors) for the ladder and ladder plus  cross-ladder kernels with the ones obtained with $N=3$. It is visible that also for the light-front wave functions, the cross-ladder
 effects are largely suppressed when color is included in the calculations. This suppression is even more clearly demonstrated in 
 Fig.~\ref{wavefunction-ratios}, where we display the ratio between light front wave functions corresponding to the L and L+CL kernels with 
 respect to $\gamma$ (left panel) and  $z$ (right panel). It is seen in the left panel that with no color factors the cross-ladder 
 contribution is above 40\% for large values of $\gamma$. However, for $N=3$ the cross-ladder effect is reduced to about $3\%$. 
 In addition, the effect of the  cross-ladder  on the tail of the wave function at large $\gamma$ turns to be proportional to the square of 
 the coupling constant, without changing its power law dependence determined by the ladder exchange. 
 In Ref.~\cite{Gigante17} it was deduced that in the asymptotic limit $\Psi^{\text{L}}_{\text{LF}}/\Psi^{\text{L+CL}}_{\text{LF}}\approx g^2_{\text{L}}/g^2_{\text{L+CL}}$. Our calculations thus show that the same relation holds when color factors are included in the model.

\begin{figure}[!hbt]
\begin{center}
\includegraphics[scale=.4]{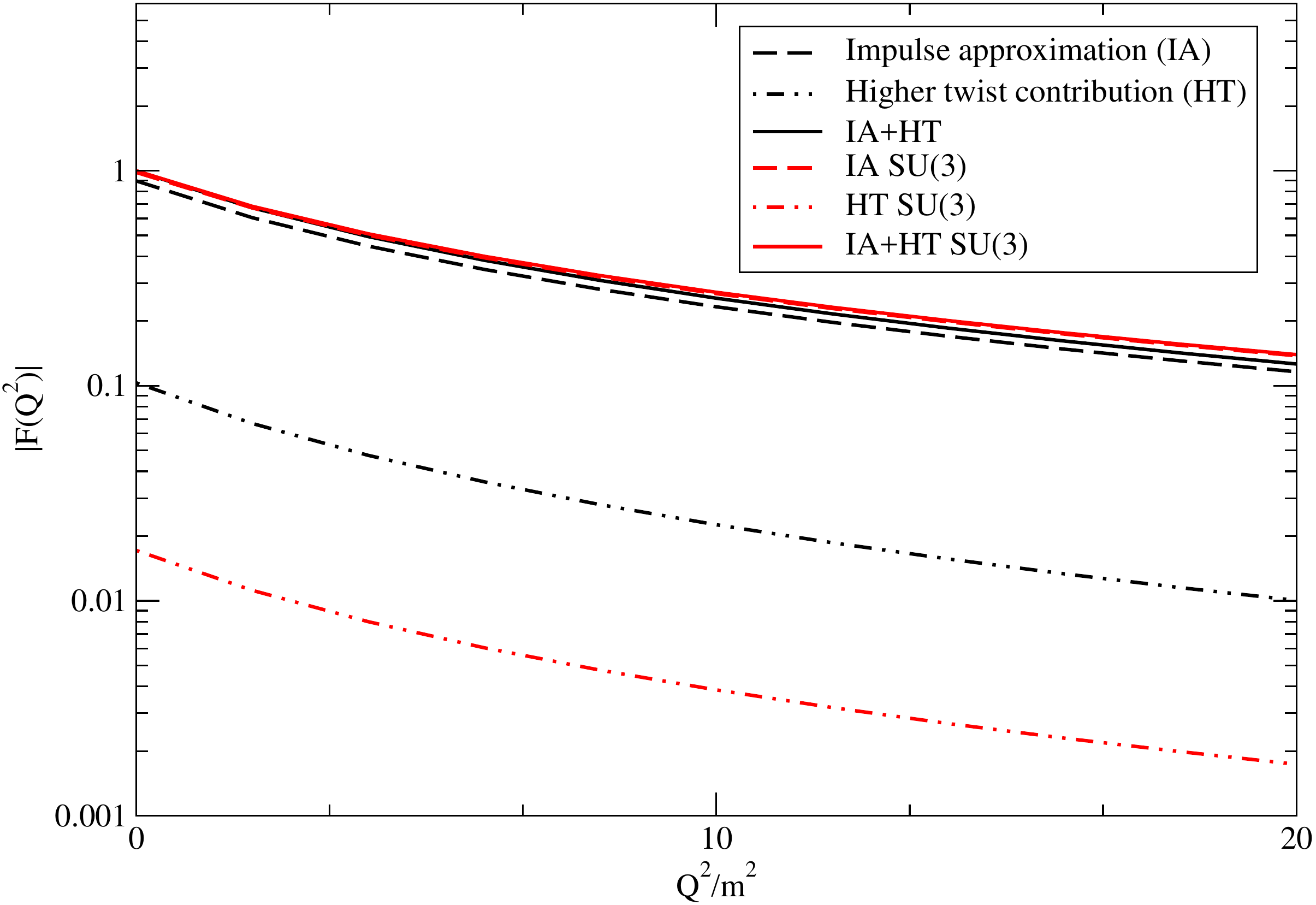}
\caption{EM form factor beyond the impulse approximation (IA), considering the contribution from the first cross graph, namely, the higher twist contribution (HT) for $\mu=0.001\,m$. Comparison with the one computed with $N=3$. In the calculations a binding energy of $B=1.0\,m$ was used.}\label{formfactor-Nc}
\end{center}
\end{figure}

Finally, we also studied the suppression of the cross-ladder effects on the elastic electromagnetic (EM) form factor, by performing calculations based on the  
formalism outlined in Ref.~\cite{Gigante17}. The calculated impulse approximation and two-body current contributions to the EM form factor as functions of the  
momentum transfer $Q^2=-q_\mu q^\mu$ are shown in Fig.~\ref{formfactor-Nc}, for the binding energy $B=1.0\,m$, and the exchanged mass $\mu=0.001\,m$. For sake of simplicity it is not explicitly presented, but it was checked that, as expected, the charge radius or the slope of the EM form factor at $Q^2=0$ is larger for $\mu=0.001\,m$ than for $\mu=0.5\,m$ due to the longer range interaction for the smaller exchanged mass. In the figure, the results obtained with
$N=3$  are compared with the ones computed without color factors. We confirmed that the inclusion of the color degree of freedom also leads to a large suppression of the two-body current contribution to the EM factor. For example for $Q^2\approx 0$ the ratio is about $13\%$  without any color factor and it is then reduced to $1\%$ in the case of $N=3$.
\vspace{1cm}
%

\section{Summary and Outlook \label{Sec:Summary and Outlook}}
We studied the cross-ladder kernel contribution in a prototype of the flavor-nonsinglet meson system composed by a scalar-quark and a scalar-antiquark with equal masses. It was found that the inclusion of the color factors in the irreducible kernel of the Bethe-Salpeter equation suppresses the huge effect of the non-planar Feynman diagram in all the studied observables: coupling constant for a given binding energy, valence light front wave-function and elastic form factor, where we found that the two-body current contribution from the cross-ladder is suppressed.

The color factors can also be easily incorporated in the calculations of two-fermion systems, such as the ones performed in Refs.~\cite{Carbonell10,Wayne}. 
In this way a more realistic description of mesons, such as the pion, could be obtained, which is the task that we hope to pursue in future studies. 

Our study supports, in a simple model, the rainbow-ladder truncation used for calculations of hadron observables in QCD. As a prospect for the next studies, 
we should consider self-energies  by means of the K\"allen-Lehmann spectral representation, which will put our model calculations in a even more 
realistic scenario.
The unequal mass case with $m_1 \to \infty$ while $m_2$ fixed will also be subject of the forthcoming study to understand how the one-body limit of the BSE is recovered when introducing the color degrees of freedom.

\section*{Acknowledgments}
J.H.A.N. acknowledges the support of the grant \#2014/19094-8 and \#2017/14695-1 from S\~ao Paulo Research Foundation (FAPESP) and the grant from the APS/SBF interchange program of 2015.
C.-R. Ji's work was supported by the U.S. Department of Energy (Grant No. DE-FG02-03ER41260) and the Conselho Nacional de Desenvolvimento Científico e Tecnológico (CNPq) of Brazil (401322/2014-9, 308025/2015-6, 308486/2015-3). E.Y.~thanks for the financial support of the grant \#2016/25143-7 from FAPESP. TF
thanks the FAPESP Thematic Project grant No. 17/05660-0.
This work is a part of the project INCT-FNA Proc. No. 464898/2014-5.
One of the authors (J.H.A.N) is grateful for the warm hospitality of the Prof. C.-R. Ji group at North Carolina State University.

\end{document}